# Return to Bali

## (*machine learning for ethnobotany in context*)

Marc Böhlen
Department of Art,
Emerging Practices in
Computational Media
University at Buffalo
*Buffalo, USA*
marcbohlen@protonmail.com

Wawan Sujarwo
Ethnobiology Research Group
Research Center for Biology
Indonesian Institute of Sciences
*Cibinong, West Java, Indonesia*

wawan.sujarwo@lipi.go.id

*Abstract*—This paper gives an overview of the project *Return to Bali* that seeks to create a living dataset of ethnobotanically significant flora on the island of Bali and new methods through which underrepresented forms of knowledge can be documented, shared and made compatible within the logics of machine learning while considering practical approaches to benefit multiple stakeholders and preventing unintended harm.

*Keywords—ethnobotany, machine learning, knowledge representation, local ecological knowledge.*

## I. INTRODUCTION – BAYUNG GEDE

In the 1930s, social anthropologists Margaret Mead and Gregory Bateson performed extensive field research to study and understand the culture of Central Balinese peoples in Bayung Gede [2]. The film *Trance and Dance in Bali* (1952) and the book *Balinese Character: A Photographic Analysis* (1942) emerged as both deep inquiries into aspects of life in the village of Bayung Gede as well as a new approach to documentation at large. In over 700 photographs with accompanying explanatory comments and interpretative summaries pointing out connections across the visual media, the book and film show how Mead and Bateson attempted and largely succeeded in creating a multimedia experiment ahead of its time. The researchers understood that the synergies across these distinct techniques could capture insights that any single method performed in isolation could not. They combined photography, film, and note taking into a comprehensive 'reality capture' machine, with photography targeting visual focus, with film enabling extended observation over time, and handwritten notes applied to capturing thoughts and ideas. In fact, the lasting contribution of the project today lies primarily in its contribution to social science methodology as it launched the field of visual anthropology [15].

## II. SECOND ORDER CYBERNETICS HANGOVER

Mead and Bateson's work are of interest to this project not only for their experimentation in novel multi-media, but also for the intellectual territory they sought to apply then novel technologies in the first place. Both Mead and Bateson played significant roles in contributing to the agenda of Second Order Cybernetics [11]. While the differences between the two classes of cybernetics are far less crisp than the ordinal numbering suggests, Second Order Cybernetics did place a unique focus on the social dimensions of information technologies, seeking explicitly holistic approaches to the application of control technologies toward better lives for multitudes, and applying adaptive systems to organizational and political contexts, laying the conceptual foundations for the extension of the scope of information technologies and artificial intelligence into the realm of well-being at large [14]. The success of engineering-focused First Order Cybernetics is legend; Artificial Intelligence and Robotics define key parts of the 21st century technical landscape. The failure of Second Order Cybernetics to produce similarly effective procedures by which to implement its bold ideals leave to this day a gaping hole in the efforts to integrate Artificial Intelligence into socially relevant and responsible activities.

## III. MACHINE LEARNING DEFICIENCIES

The fact that Artificial Intelligence and Machine Learning struggle with structural problems of unfairness and bias can be traced in part to this early schism. While uneven fairness has been recognized and debated in Economics for a long time [25], researchers have only recently responded with concrete proposals by which to address some of the problems [3], [10], [20], including the role of big data [1], specifically in the context of the ethics of machine learning. In this context, algorithm design meant to mitigate unintended negative fallout from predictive policing [16] or facial recognition [17] as well as the attempt to promote fairness [7], [13] have all received attention, yet some voices have – in our view correctly - pointed out that algorithmic interventions alone are not sufficient to generate effective fairness across a project pipeline [7], [23].

Machine learning data sets are often biased due to the preferences and needs of the researchers who compile the materials. Furthermore, supervised learning systems can only perform within the opportunities and limitations of the data on which they have been trained on. Fundamentally, a dataset is a sample derived from a larger population, and the sample must have the same statistical features as the population in order to serve as a viable representation of the larger entity. If the sample does not properly represent the population, any conclusions about the larger population drawn from the sample will be by definition invalid. Moreover, failing to represent a target population does not present itself as a binary, but rather as a sliding condition to which incorrect, partial, non-representative or low-quality data can negatively contribute in various ways.

| Name | Source | Description | Collected by |
|---|---|---|---|
| ImageNet (since 2009) | Internet | 14 million+ URLs of images, hand-annotated. 20'000+ categories. Uses a variant of WordNet | Crowdsourced labels via Mechanical Turk workers. |
| SVHN (2011) | Street View images | 600,000+ digit and number images in altering contexts. | Automated algorithms and Mechanical Turk workers to transcribe digits. |
| LSUN (since 2015) | Internet | Iteratively collected set of 59 million images for 20 object categories | Trained experts, crowdsourced labels via Mechanical Turk workers + automated labelling. |
| CoCo (since 2015) | Internet | Image recognition in context. 330K images (>200K labeled) and 1.5 million object instances with 80 object categories. | Crowdsourced labels via Mechanical Turk workers. |
| CelebA (since 2015) | Internet | Face attributes dataset with more than 200K celebrity images, each with 40 attributes. | Crowdsourced labels via paid participants |
| CORe50 (2017) | Custom made (Kinect sensor) | 164,866 128×128 images of 50 objects in 10 categories, 11 temporally coherent sessions; camera point-of-view of operator | Trained experts |

*Table 1. Popular image repositories for machine learning classification*

Table 1 shows an overview of some of the popular image data sets used in machine learning. For our project, it is important to note that in the last decade, image materials for machine learning have been mostly collected from general Internet resources. Moreover, these images are often labeled by anonymous human workers and automated evaluation procedures. Mechanical Turk has become a popular approach for many research teams to outsource the labelling process to remote human workers. While the outsourcing of the data labeling effort lowers project costs, unintended slippages and information loss can occur, specifically when image creators and labelers do not share the same cultural background, as a recent inquiry into the production of the *CELEBA* dataset has shown [6]. Furthermore, the current distribution of images used in computer vision research is unevenly distributed in their geographical origin, with a large majority of images from the Open Images data set, for example, sourced from North America and Europe [18]. For all these reasons, our project places a premium on the data collection strategies, including the social and economic dynamics that accompany the process of theme identification and subsequent data collection in the field.

## IV. BALIPLOITATION

Mead and Bateson's research serves as a conceptual springboard for our project not only due to its influence on Second Order Cybernetics, but also for Bateson's foundational contribution to the very concept of postmodern ecological consciousness [8]. We are also fully cognizant of the fact that in the wake of Mead and Bateson's publication, the island of Bali has been subjected to a wave of exploitative activities. The 1932 documentary *Virgins of Bali,* sporting scenes of topless Balinese women while nudity of white women was banned in Hollywood, "almost single-handedly made Bali into a popular spot for tourists" [9].

While tourism did in fact allow Bali to increase economic opportunities for many local inhabitants, the model of an almost total reliance on tourism as a main source of income has proven fragile. This model is highly sensitive to global political forces that would otherwise not affect the tropical island. For example, Bali continues to suffer enormous damage due to the disruption of global travel and tourism in the wake of a global epidemic only recently relegated to the realm of science fiction.

## V. RETURN TO BALI

Broadly, *Return to Bali* asks the following question: How can we represent forms of knowledge that have to date not been represented in machine learning? Currently, the critical discourse on representation in machine learning systems in America and Europe focuses on the misuse of representation, particularly when existing data or algorithms are inadequate or misused for nefarious purposes, surveillance being the best documented example.

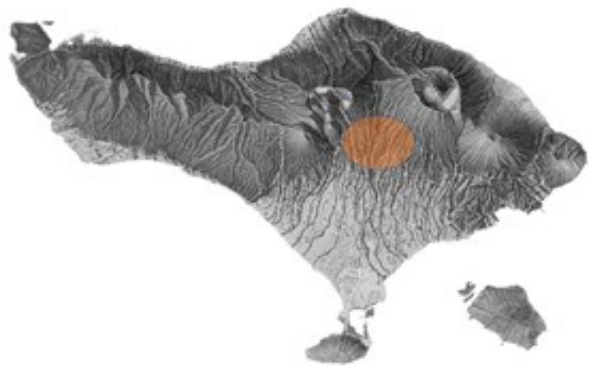

Figure 1. The island of Bali with the field study site in Central Bali.

To be clear: In no way does this project take the many dangers of compromised privacy lightly, and yet we present a different focus in this study. From the vantage point of emerging economies – to the degree that this project can speak to such vast territory - *a completely different concern exists in the condition of not being represented at all*, in not having a place in the new world of digital information and machine learning processing, as other researchers have also observed [18]. What is not formally represented in this differently structured world can have no defensible rights in the knowledge society of the 21$^{st}$ century.

Indonesia is one example of this condition in which the desire to participate in the 21$^{st}$ century global knowledge economy clashes with privacy management concerns the global north has focused on. And so the situation in our case offers double trouble as well as double opportunities: to both address the right of representation in the 21$^{st}$ century knowledge economy, and to respond at the same time to the challenges of situation specific bias and unfairness within the machine learning framework that enables this economy.

In order to practically address these intertwined challenges, we select a specific question, place and context in which to apply the study. Our topic is that of ethnobotany, the study of how people and societies use and interact with plant life at large. We focus our study on the Island of Bali, specifically in Central Bali (Fig. 1), where Mead and Bateson conducted their first multi-media studies and we ask specifically how machine learning might assist in counteracting the documented decline of local ecological knowledge of plants and their uses in everyday life amongst the current generation of Balinese youth. In particular, knowledge of food and nutraceutical plants has declined amongst younger Balinese in traditional villages [20].

One might argue that this project is too broad and vague yet our goal is foremost the design of an alternate way of performing transdisciplinary A.I.; an approach that confronts A.I development with constraints it might otherwise be shielded from; moving all parts of the project forward with equal concern [4]. The following sections describe the project in broad strokes only. They elaborate on the procedures we have put into place in order to achieve our goal, including the selection of categories in our collection, the methods deployed in creating the collection, the knowledge of the project participants and how the lessons from Mead, Bateson and machine learning ethics inform the concept and practicalities of assembling this A.I. project in the first place.

## VI. COLLABORATION ACROSS DISCIPLINES AND ECONOMIC GRADIENTS

*Return to Bali* began with lengthy electronic exchanges between members of the core research team. This introductory period was followed by joint field work performed during the months of February and March 2020 in Central Bali during which time the bulk of the first dataset was collected. Subsequent additions to the first dataset were performed by members of the local team in consultation with the remotely operating research team. These activities produced a data collection and control regime that allows us to perform quality control over the entire data collection and ingestion process as well as to maintain social bonds that sustain a project with distributed participants. By integrating digital payments via a social media application popular with our local team members, we can remunerate the local data harvesters as they perform their work in the field. Furthermore we make use of the fluctuations in currency markets that often negatively impact emerging economies (1$US trades for approximately 16'000 Indonesian Rupiah) to support our local partners by issuing fixed payments in the strong currency from the team member in the global north. The result of these investments are a research effort in which all team members profit in different ways.

## VII. REPRESENTING LOCAL ECOLOGICAL KNOWLEDGE

Conceptually, *Return to Bali's* most difficult challenge is the question of how to represent local ecological knowledge in the first place. Many aspects of local ecological knowledge are informal and experiential, grounded in personal and community experiences and experiments collected in some cases across generations. Ethnobotany seeks to do justice to these forms of knowing while acknowledging the significance of state of the art botany. Our project attempts to navigate these sometimes divergent vectors with ecological knowledge sourced from several individuals in Central Bali combined with scientific grounding in ethnobotany research, under the influence of machine learning as a helper function.

A key project participant is Made Darmaja. Darmaja maintains a forest garden in Kerta village with dozens of plants and fruits native to Southeast Asia and several examples of plants indigenous to Malesian region, including Bali, such as carambola (*Averrhoa carambola*), durian (*Durio zibethinus*), Indonesian cinnamon (*Cinnamomum burmannii*), mangosteen (*Garcinia mangostana*), and snake fruit (*Salacca zalacca*). He has developed an informal, holistic approach to caring for this plot of land, keeping pests at bay while allowing for bees to proliferate and pollinate the fruit trees, practicing aquaculture in a small-pond, especially for catfish, and raising chickens and cattle on the land, performing an adaptive and highly improvised form of integrated farming (Fig. 2) without any formal training in the field. Other members of the local data harvesting team, Gusti Sutarjana and his family, own a plot of land in Bukian village that has been in family possession for over seven generations. This mature forest plot has cacao (*Theobroma cacao*), taro (*Colocasia esculenta*), patchouli (*Pogostemon cablin*), sugar palm (*Arenga pinnata*) and many plant species of relevance to various local culinary and medicinal practices (Fig. 3).

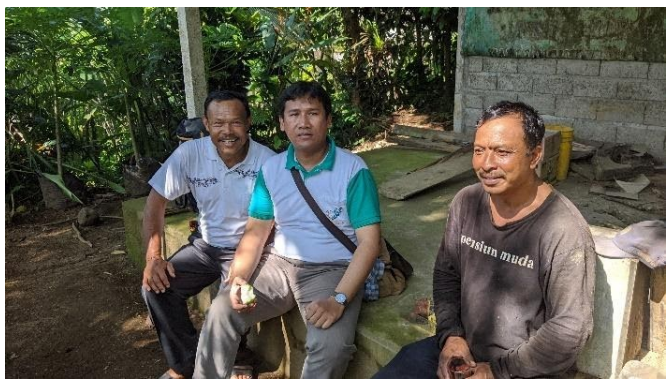

Figure 2. Peneng (left), Wawan (center) and Darmaja (right) in Kerta Village

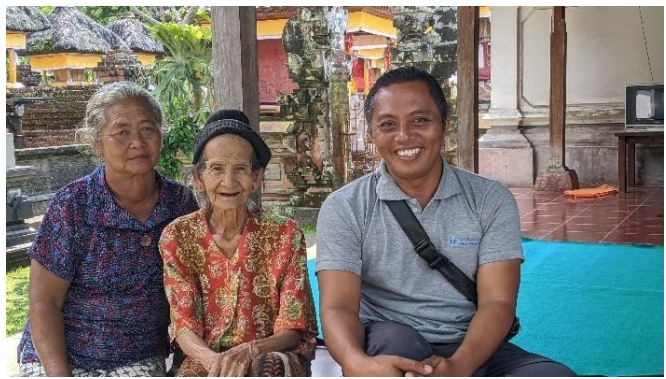

Figure 3. Jero Ade (left), Jero Nini (center) and Gusti (right) in Bukian Village

## VIII. Machine learning in ethnobotany

Traditionally, ethnobotany research has relied on documentation of villagers' knowledge through in-person interviews collected with time-based electronic media (audio/film/video) and or hand-written notes that are then manually evaluated and contextualized by ethnobotanists. To date, machine learning has not been applied to ethnobotany field studies. Our work provides a first preliminary system that integrates established practices into a novel machine learning compatible framework that supports state of the art plant classification. Specifically, we introduce a new way to capture local ecological knowledge in the field with mobile phones in order to prepare the data for ingestion into machine learning classification (Fig. 4). The approach uses the audio utterances in video streams as labels for images and preprocesses the combined information for ingestion into standard convolutional neural networks that comprise the current default approach to automated image classification as described in a companion document [5]. Moreover the video feeds can be post-annotated such that experts not present during data collection in the field can assist and, if necessary correct, the labeling effort, enabling new forms of distributed workflows that can make better use of remote or distributed expertise (Fig. 5). Following this field data ingestion step we perform quality control on the data and train state of the art supervised image classifiers on the data set. Parallel to the image processing, we use speech-to-text machine technology to facilitate audio transcription of in-person interviews. All these field video processing steps are combined in an open source software package, *Catch & Release*[1].

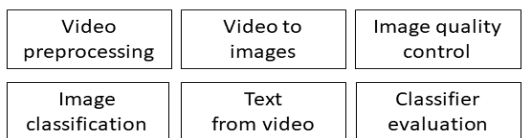

Figure 4. Elements of the video processing software *Catch & Release*.

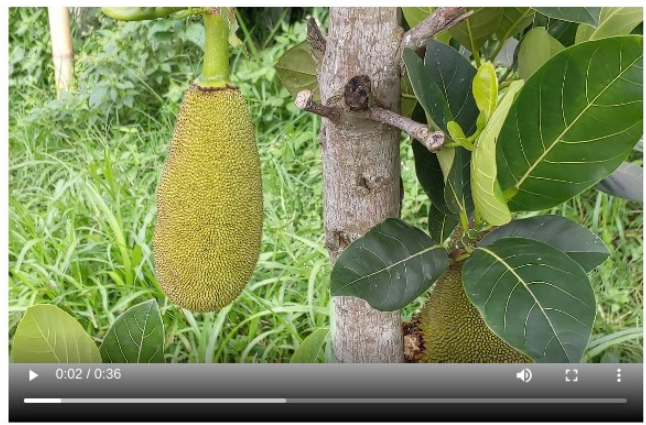

Figure 5. Audio annotation in the user interface of *Catch & Release*

Machine learning databases are often one shot representations of a particular theme. Ethnobotany not only requires – as does machine learning - copious amounts of data, but also varied examples of any particular category. For example, we wish to represent the flora in the wet as well as the dry season as not all plants flower in the same season. Moreover, some plants have specific uses at one stage of growth, and very different uses at later stages of growth. In fact, most of the plants in our collection have a multiplicity of uses. For example, sugar palm is a multipurpose tree that is considered a cultural keystone species in Bali. Young leaves are used for making cigarette paper and salad; the sap of male inflorescences is tapped to produce fresh drink, wine, vinegar, and sugar palm flowers are a source of nutrition for bees producing honey. The stem of the sugar palm is used as a building material while the inner stems serve as a staple food, boiled fruits can be eaten, fibers from the leaf sheaths are used to make rope, brooms, and roofing material, and a decoction of the roots is used to treat urolithiasis [22]. Also, the young shoots of bamboo petung (*Dendrocalamus asper*) are used as vegetables while the mature culm is used as a building material. Likewise the young fruit of papaya (*Carica papaya*) is a desirable vegetable while the ripe fruit is only consumed as a fruit.

---

[1] https://github.com/realtechsupport/c-plus-r

The bali-26 dataset in its current state does not adequately address all these complexities. As such, our representation suffers both from an unusual form of representation bias as well as oversimplification [23]. While our current collection covers only the wet season, it does combine fruits, leaves and trees of the categories where such examples were available. Yet this compilation only scratches the surface of ethnobotanical territory. For example, we have yet to find a way to represent the results of processing of plants into culinary products or the important role Balinese devotional practices play on the use of these plants.

An important aspect of our project, and one that mitigates to some degree these deficiencies, is its ability to continue to expand the data collection over time. This feature is enabled both through the arrangement of our technology as well as through our collaboration pipeline (Fig 6). Our team made use of this opportunity when it became clear that we had erroneously collected multiple species of bamboo during one of the field trips.

| 1 Select plant species | 2 Contact local team | 3 Select field site | 4 Collect video in the field |
|---|---|---|---|
| 5 Send video to researchers | 6 Expert assessment | 7 Remunerate local team | 8 Video to images |
| 9 Images to training data | 10 Retrain classifiers | 11 Evaluate results | 12 Update collection |

Figure 6: Steps put in place to add new data to the dataset.

The local team then was tasked with locating examples of bamboo petung, collecting high definition videos of several examples in the wild (Fig. 7), and posting them via mobile phones to our remote data server in Switzerland with strong digital rights protections from where the research team collected and evaluated the videos and then remunerated the data harvesters accordingly (Fig. 8).

We describe in a companion document how our video-to-labeled-images system works, how we trained several machine learning classifiers on the resultant data, and how the classifiers performed in several different tests [5]. The result of these operations is a curated image collection of 26 ethnobotanically relevant plants and fruits of Central Bali represented in over 50'000 individual images (Fig 9), together with several trained classifier models as well as the *Catch & Release* software interface with which one can engage with the materials and adapt the framework to other research endeavors.

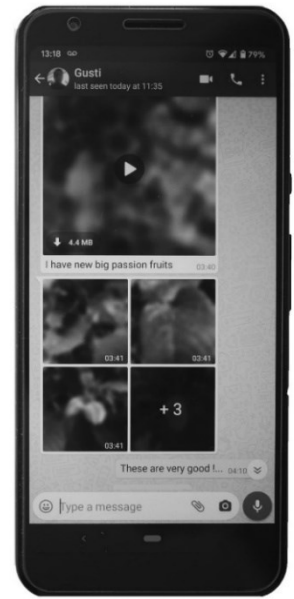

Figure 7. Field videos sent from local data harvesters to the research team for evaluation.

Applying machine learning to ethnobotany also generates new research questions for machine learning proper. While the challenge of representing the interwoven threads of knowledge, experience and best practices contained within local ecological knowledge remains unsolved, other issues are more accessible, such as multi-label classification [24] designed to associate any one image with multiple descriptors. Yet not everything possible is advisable. With the availability of global location information in mobile media recordings, image sources can easily be geotagged. Yet this powerful extension does not always scale well, and the potential for misuse has been discussed in various contexts [12].

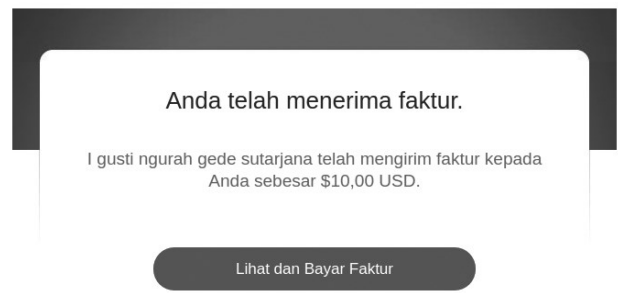

Figure 8. Confirmation of electronic payment to local data collectors.

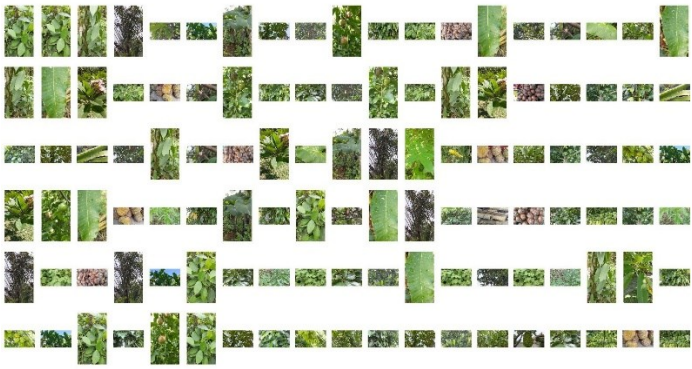

Figure 9. Samples from the current version of the bali-26 data set.[2]

## IX. Learning from the return to Bali

In order to ensure that larger parts of the population benefit from documenting their own local ecological knowledge we have to define better methods of returning the output to the general public. Possibly integrating some parts of the project into curricular activities could help as well as providing to local communities more agency regarding project scope and limitations.

There can be no doubt that the system as it currently stands falls short of our original intentions. And yet the connection between the individual parts and the construction philosophy that drives it will hopefully not only assist ethnobotany as it positions itself to harness machine learning techniques for its own needs, but identify one possible path forward by which underrepresented forms of knowing can responsibly be integrated into the fold of machine learning opportunities while preserving the tenets of well-being in general.

Beyond the confines of this project we hope that other researchers will make use of the software we have created to capture under-represented knowledge in many more contexts, and that students of machine learning will find more, and more varied, datasets not recycled from the internet with which to train their skills, and strengthen their appreciation and critique of machine learning systems in the future.

## Acknowledgments

We would like to express our gratitude to Gusti Sutarjana, his family, and Made Darmaja for sharing their knowledge, hospitality, and assistance. We also express our appreciation to Nyoman Peneng for his assistance during fieldwork.

## References

[1] S. Barocas, A. Selbst. 2016. "Big data's disparate impact". California Law Review, 104.
[2] G. Bateson, M. Mead, "Balinese character; a photographic analysis", New York Academy of Sciences, 1942.
[3] S. Bird, S. Barocas, K. Crawford, F. Diaz, H. Wallach. 2016. "Exploring or Exploiting? Social and Ethical Implications of Autonomous Experimentation in AI" (October 2). Workshop on Fairness, Accountability, and Transparency in Machine Learning.
[4] M. Böhlen. 2020. A.I. has a rocket problem. The Startup. Medium. August 3. https://medium.com/swlh/ai-has-a-rocket-problem-6949c6ed51e8
[5] M. Böhlen, W. Sujarwo. 2020. "Machine Learning in Ethnobotany". IEEE International Conference on Systems, Man, and Cybernetics. Toronto, Canada.
[6] M. Böhlen, V. Chandola, A. Salunkhe. 2017. "Server, server in the cloud. Who is the fairest in the crowd?" https://arxiv.org/abs/1711.08801
[7] A. Bower, S. Kitchen, L. Niss, M. Strauss, A. Vargas, S. Venkatasubramanian. 2017. "Fair Pipelines". arXiv:1707.00391v1
[8] A. Chaney. 2017. "Runaway. Gregory Bateson, the Double Bind, and the Rise of Ecological Consciousness". Chapel Hill. The University of North Carolina Press.
[9] T. Doherty. 1999. "Pre-Code Hollywood: Sex, Immorality, and Insurrection in American Cinema, 1930–1934". New York: Columbia University Press, p. 134.
[10] C. Dwork, M. Hardt, T. Pitassi, O. Reingold, R. Zemel. 2011. "Fairness Through Awareness". CoRR abs/1104.3913. http://arxiv.org/abs/1104.3913
[11] R. Glanville, 2002. "Second order cybernetics". In F. Parra-Luna (Ed.), Systems science and cybernetics. In Encyclopedia of life support systems (EOLSS).
[12] M. Gregory. 2019. "Stop Blaming Instagram for Ruining the Great Outdoors". Vice. July 26.
[13] M. Hardt, E. Price, N. Srebro. 2016. "Equality of Opportunity in Supervised Learning". CoRR abs/1610.02413 https://arxiv.org/abs/1610.02413
[14] IEEE. "Well-being". 2020. The IEEE Global Initiative on Ethics of Autonomous and Intelligent Systems. https://standards.ieee.org/content/dam/ieee-standards/standards/web/documents/other/ead1e_well_being.pdf
[15] I. Jacknis. 1988. "Margaret Mead and Gregory Bateson in Bali: Their Use of Photography and Film". Cultural Anthropology, Vol. 3, No. 2, pp. 160-177.
[16] K. Lum, W. Isaac. 2016. "To predict and serve?" Significance 13(5): pp.14–19.
[17] P. Phillips, F. Jiang, A. Narvekar, J. Ayyad, A. O'Toole. 2011. "An other-race effect for face recognition algorithms". ACM Transactions on Applied Perception (TAP) 8(2):14.
[18] S. Shankar, Y. Halpern, E. Breck, J. Atwood, J. Wilson, D. Sculley. 2017. "No Classification without Representation: Assessing Geodiversity Issues in Open Data Sets for the Developing World". NIPS 2017 workshop: Machine Learning for the Developing World.
[19] M. Skirpan, M. Gorelick. 2017. "The Authority of Fair in Machine Learning". KDD'17 FATML Workshop, Nova Scotia, CANADA.
[20] W. Sujarwo, I. Arinasa, F. Salomone, G. Caneva, S. Fattorini. 2014. "Cultural erosion of Balinese indigenous knowledge of food and nutraceutical plants". Economic Botany 68, no. 4, 2014, pp. 426–437.
[21] W. Sujarwo, and G. Caneva. 2016. "Using quantitative indices to evaluate the cultural importance of food and nutraceutical plants: Comparative data from the island of Bali (Indonesia)". Journal of Cultural Heritage, no. 18, 2016, pp. 342–348.
[22] W. Sujarwo, G. Caneva, and V. Zuccarello. 2019. "Bio-cultural traits and cultural keystone species, a combined approach: an example of application about plants used for food and nutraceutical purposes in Aga villages in Bali, Indonesia". Human Ecology 47, no. 6, 2019, pp. 917-929.
[23] H. Suresh, J. Guttag. 2019. "A Framework for Understanding Unintended Consequences of Machine Learning". arXiv:1901.10002v3
[24] J. Wehrmann et al. 2018. "Hierarchical Multi-Label Classification Networks". Proceedings of the 35th International Conference on Machine Learning, in PMLR 80: pp. 5075-5084.
[25] P. Young. 1995. "Equity. In Theory and Practice". Princeton University Press.
[26] J. Zou, L. Schiebinger. 2018. "AI can be sexist and racist - it's time to make it fair". Nature 559, pp. 324-326.

---

[2] available at: https://filedn.com/lqzjnYhpY3yQ7BdfTulG1yY/bali-26.zip